\documentclass[prd,aps,showpacs,nofootinbib,eqsecnum,onecolumn,superscriptaddress,amssymb]{revtex4}
\usepackage{amsmath}
\usepackage{amssymb}
\usepackage{amsfonts}
\usepackage{graphicx,bm}
\usepackage{dcolumn}
\usepackage[colorlinks=true]{hyperref}
\usepackage{color,amsxtra}
\usepackage{epsf}
\usepackage{enumerate}
\usepackage{hhline}
\usepackage{array}
\usepackage{tabularx}
\usepackage{subfigure}
\usepackage{fancyhdr}
\usepackage{mathrsfs}

\newcommand{\be}{\begin{equation}}
\newcommand{\ee}{\end{equation}}
\newcommand{\bea}{\begin{eqnarray}}
\newcommand{\eea}{\end{eqnarray}}
\newcommand{\beaa}{\begin{eqnarray*}}
\newcommand{\eeaa}{\end{eqnarray*}}

\def\be{\begin{equation}}
\def\ee{\end{equation}}
\def\bea{\begin{eqnarray}}
\def\eea{\end{eqnarray}}

\begin{document}


\title{Exponential gravity with logarithmic corrections in the presence of axion dark matter}

\author{Sergei D. Odintsov}
\email{odintsov@ice.csic.es} 
 \affiliation{Instituci\'o Catalana de Recerca i Estudis Avan\c{c}ats (ICREA),
Passeig Luis Companys, 23, 08010 Barcelona, Spain}
\affiliation{Institut de Ci\`{e}ncies de l'Espai,
ICE/CSIC-IEEC, Campus UAB, Carrer de Can Magrans s/n, 08193 Bellaterra (Barcelona),
Spain}
 \author{Diego~S\'aez-Chill\'on~G\'omez}
\email{diego.saez@uva.es}
\affiliation{Department of Theoretical, Atomic and Optical
Physics, Campus Miguel Delibes, \\ University of Valladolid UVA, Paseo Bel\'en, 7, 47011
Valladolid, Spain}
\affiliation{Department of Physics, Universidade Federal do Cear\'a (UFC), Campus do Pici, Fortaleza - CE, C.P. 6030, 60455-760 - Brazil}
\author{German~S.~Sharov}
 \email{sharov.gs@tversu.ru}
 \affiliation{Tver state university, Sadovyj per. 35, 170002 Tver, Russia}
 \affiliation{International Laboratory for Theoretical Cosmology,
Tomsk State University of Control Systems and Radioelectronics (TUSUR), 634050 Tomsk,
Russia}

%

%
%
\begin{abstract}

An exponential modified gravity with additional logarithmic corrections is considered with the presence of an axion-like scalar field in the role of dark matter. Axion fields are thought to become important at late-times when the axion-like scalar field oscillates around its vacuum expectation value, mimicking dark matter behaviour. The model is compared with the usual pressureless fluid description of dark matter. Both models are tested with observational data including some of the latest sources, providing similar fits in comparison with the $\Lambda$CDM model. Despite results are not statistically relevant to rule out any model, the number of free parameters still favours $\Lambda$CDM model, as shown by computing the goodness of the fits.

\end{abstract}
\pacs{04.50.Kd, 98.80.-k, 95.36.+x}
\maketitle
%
%
%

\section{Introduction}

Cosmology is currently under great challenges that are pushing the research forward of its limits. In this sense, General Relativity (GR) requires the presence of new matter components or some appropriate modifications in order to satisfy the observational data. Among current cosmology's issues, dark matter, dark energy and inflation stand about over the rest \cite{reviews1}. Among modified theories of gravity one can note a special role of the so-called $F(R)$ modified gravity, particularly when dealing with cosmological models. The essential of this type of GR extension lies on considering Lagrangians non-trivially depending on the Ricci scalar $R$. These models show to provide a successful description of both the inflationary era as the late-time epoch when the universe expansion accelerates  \cite{reviews1,reviews2}. For instance, the so-called  $R^2$ or Starobinsky inflation \cite{Starobinsky1980} is one of the most promising inflationary models, since satisfies the corresponding Cosmic Microwave Background (CMB) constraints, as provided by Planck collaboration  \cite{Planck13,Planck18}. Also late-time acceleration can be well described in the framework of $F(R)$ gravities, since almost for every cosmological solution, a particular $F(R)$ gravitational action might be obtained \cite{Capozziello2002,unifying}. In addition, deviations from GR predictions at local scales might become small enough such that local constraints, as in the Solar System, can be also satisfy in $F(R)$ gravities  \cite{Hu:2007nk,Nojiri:2007cq,Linder2009}. Then, cosmological scenarios within $F(R)$ gravity might describe the entire cosmological evolution, satisfying theoretical constraints as well as observational limitations
\cite{Appleby:2007vb,Ravi:2023nsn,Oliveros:2023ewl} for both the early and late-time acceleration eras.\\

Among $F(R)$ models a particular class of exponential gravities that contain terms as $e^{-bR}$ in the action can successfully describe the cosmological evolution. In such type of models, the parameter $b$ manages the scale or the cosmological epoch when the exponential plays an important role \cite{Linder2009,nostriexp}. Moreover, the exponential term can provide a fast transition when the curvature scalar reaches the scale $R\sim b^{-1}$ while such a term becomes negligible for $b R>>1$, which occurs as far as the curvature scalar is large enough (at the early epoch) and/or the parameter $b$ is also large. Hence, for a type of exponential gravity whose Lagrangian has the form  $R-2\Lambda\big(1-e^{-bR}\big)$, the model mimics the $\Lambda$CDM model and provides a successful cosmological evolution in comparison with other models and with observational constraints \cite{YangLeeLG2010,OdintsovSGS:2017,OdintsovSGStens:2021}. In addition, exponential gravities may include also terms that act effectively during inflation and thereby the models can reproduce the whole cosmological evolution \cite{nostriexp}. In particular, such inflationary terms may be constructed based on $R^2$ inflation with an appropriate factor
that makes the contribution to become important at early times (for a large curvature scalar) but is suppressed at late-times \cite{nostriexp,OdintsovSGS:2017}. Also, such exponential models might be considered as effective corrections motivated by quantum gravity effects
\cite{Buchbinder:1992rb,Cognola:2005de}. In particular, these corrections include also logarithmic terms, which can evolute smoothly \cite{Cognola:2005de,Nojiri:2003ni,Myrzakulov:2014hca,Elizalde:2017mrn,OdintsovOS:2017log,OdintsovSGSlog:2019}. Note that modification of exponential gravity with some extra logarithmic terms have been previously considered in the literature  \cite{OdintsovSGSlog:2019}. By testing this type of models, one finds optimistic results when comparing the cosmological evolution, including the inflationary era, primordial nucleosynthesis and late time observational data, with the $\Lambda$CDM model.\\

The aim of this paper is to analyse the viability of a class of exponential $F(R)$ gravity with some logarithmic corrections \cite{OdintsovSGSlog:2019}. To do so, two scenarios for describing dark matter are considered: the usual description of dark matter as a pressureless fluid and an axion-like particle. Note that the latter is one of the most promising candidates for dark matter. The axion field is described by a pseudo Nambu-Goldstone boson that was postulated as a possible solution to the CP-strong problem in quantum chromodynamics \cite{Peccei:2008}. In the original proposal, this pseudoscalar field describes a spontaneously broken global axial symmetry, the so-called Peccei-Quinn symmetry \cite{Peccei:1977}. Such field is coupled to the gluon field strength tensor that sets the CP violating term to zero thanks to a shift symmetry that absorbes indeed such contribution dynamically. Then, the axion field acquires mass via QCD instantons and despite first models were ruled out (they predicted an energy scale for the spontaneous symmetry breaking of the electroweak order), some more modern models assume a lower mass and weaker interactions \cite{Wilczek:1982}. In addition, other axion-like fields arise in other theories, as in string theory through compactifications \cite{GSWbook}. In cosmology, the axion might play different roles depending on the scale of the spontaneous symmetry breaking. However, by ignoring possible effects during the early universe and assuming that the spontaneous symmetry breaking occurs before inflation, the axion particle becomes underdamped and oscillates around an equation of state $w=0$ that provides the usual behaviour of dark matter  \cite{Sikivie:2006,Irastorza:2018dyq,Marsh:2015xka,Marsh:2017yvc,Caputo:2019tms,Soda:2017dsu}. Then, the cosmological evolution for such a field can be described by a real scalar field, which is indeed the way it will be followed in this paper \cite{Sikivie:2006,Marsh:2015xka,Irastorza:2018dyq}. In some previous papers
 \cite{OdintsovOik_UniAx:2019,OdintsovOik_Axion:2020,Oikonomou_Uni:2021,OikonomouFTR:2023,OdintsovSGS_Axi:2023}, $F(R)$ gravitational models have been analysed with the presence of an axion scalar field  that remains frozen for its primordial vacuum expectation value during the early time evolution, but start to oscillate when the Hubble rate becomes of the order of the axion mass. Despite the axion energy density  evolutes similarly to that for dust-like matter, some subtle differences might be found \cite{OdintsovOik_UniAx:2019,OdintsovOik_Axion:2020,Oikonomou_Uni:2021}. Along the present paper, the possible different cosmological evolution for both dark matter descriptions are considered in the framework of an exponential model with logarithmic terms (see Ref.~\cite{OdintsovSGSlog:2019}). Both cases are tested with observational data, including Type Ia supernovae (SNe Ia) from the Pantheon sample, estimations of the Hubble parameter $H(z)$ or Cosmic Chronometers (CC), data from Baryon
Acoustic Oscillations (BAO) and from the CMB. The scenarios are also compared with the standard $\Lambda$CDM model.\\

The paper is organised as follows: In section \ref{FRaxion}, the
Lagrangian and field equations for exponential $F(R)$ gravity with the presence of an axion are provided. Section \ref{Late} is devoted to express the equations as a dynamical system with the appropriate initial conditions. Results of the calculations and observational tests for the scenarios with a pressureless fluid as dark matter and with the axion
are analysed in section \ref{Results}. Finally, conclusions are made in section \ref{conclusions}.

\section{Exponential $F(R)$ gravity and axion}
\label{FRaxion}

Modified $F(R)$ gravity is described by a gravitational action that contains non-linear terms of the Ricci scalar, namely:
  \begin{equation}
  S =\int d^4x \sqrt{-g}\bigg[ \frac{F(R)}{2\kappa^2}  + {\cal L}_{m} + {\cal L}_{\phi}\bigg]
  ,\qquad{\cal L}_{\phi}^{(1)}=0,\quad\;{\cal L}_{\phi}^{(2)}=-\frac12\partial^\mu\phi\partial_\mu\phi-V(\phi)\, .
 \label{Act1}\end{equation}
Here $ {\cal L}_{m}$ is the matter Lagrangian, ${\cal L}_{\phi}$ describes the axion field, $\kappa^2=8\pi G$ with  $G$ being the Newtonian gravitational constant.  Along this paper  two different 
scenarios are considered: in the first one (without axion) dark matter is assumed to be part of  $ {\cal
L}_{m}$ as a pressureless fluid, such that ${\cal L}_{\phi}={\cal L}_{\phi}^{(1)}=0$.  In the second scenario the axion
scalar field in the term ${\cal L}_{\phi}={\cal L}_{\phi}^{(2)}$ plays the role of dark matter and  $ {\cal L}_{m}$ includes the remaining baryonic matter and radiation. In addition, the theoretical $F(R)$ gravity model explored here is given by \cite{OdintsovSGSlog:2019,OdintsovSGS_Axi:2023}:
 \begin{equation}
 F(R)=R -2\Lambda \left[ \big(1-e^{-\beta{\cal R}}\big)\Big(1-\alpha{\cal R}\log\frac{\cal R}2\Big)\right]+F_\mathrm{inf}(R)\ ,
\qquad  {\cal R}=\frac{R}{2\Lambda}\ ,
   \label{FR2}
\end{equation}
 where $\Lambda$ is a cosmological constant, $\beta$ and $\alpha$
are dimensionless constants. Whereas the term $F_\mathrm{inf}=\gamma(R) R^2$,
$\gamma(R)=\gamma_0+\gamma_1\log{R}$ might drive the inflationary epoch at early times but is assumed to become negligible at late-times, such that can be removed. For the purposes of the paper, a flat Friedmann-Lema\^itre-Robertson-Walker metric is considered:
 \be
 ds^2 = -dt^2 + a^2(t)\,\delta_{ij}dx^idx^j\ .
 \label{FLRW} \ee
 Then, the FLRW equations are obtained by varying the action (\ref{Act1}) with respect to the metric, leading \cite{OdintsovSGSlog:2019}:
 \begin{eqnarray}
 3 H^2F_R&=&\frac{RF_R-F}{2}-3H\dot{F}_R+\kappa^2\rho_\mathrm{tot}\ ,
 \qquad\rho_\mathrm{tot}=\rho+\rho_a\,,\label{eqn1} \\
-2\dot{H}F_R&=&\ddot{F}_R-H\dot{F}_R +\kappa^2(\rho+p+\rho_a+p_a)\ , \qquad
p_\mathrm{tot}=p+p_a\,,\label{eqn2}
 \end{eqnarray}
where $F_R\equiv \frac{dF(R)}{dR}$,  the dot denotes differentiation with respect to the cosmic
time $t$, $H=\dot a/a$ is the Hubble parameter, $\rho$ is the energy density and $p$ is the pressure of matter (including all the species of the Universe),
while $\rho_a$ and $p_a$ are the corresponding densities for the axion field. The continuity equation for the energy-momentum tensor can be easily obtained:
 \begin{equation}
 \dot\rho=-3H(\rho+p)\ .
  \label{cont}\end{equation}
In the first scenario (without axion) the energy density $\rho$ also includes dark matter while $\rho_a=p_a=0$ (together with ${\cal L}_{\phi}=0$) in the equations
(\ref{eqn1}), (\ref{eqn2}). In the second scenario the axion energy density and pressure are:
\begin{equation}
 \rho_a=\frac{1}{2}\dot{\phi}^2+V(\phi)\ ,\qquad p_a=\frac{1}{2}\dot{\phi}^2-V(\phi)\ ,
  \label{rhoa}
\end{equation}
 which play the role of dark matter in Eqs.~(\ref{eqn1}), (\ref{eqn2}). In addition, the equation for $\phi$ can be obtained by varying the action (\ref{Act1}) with respect to the scalar field, which together with the FLRW metric (\ref{FLRW}) provides the equation for the scalar field:
 \begin{equation}
\ddot{\phi}+3H\dot{\phi}+V'(\phi)=0\ .
 \label{eqphi}\end{equation}
Note that here the following potential, associated to the axion field, is considered
\cite{OdintsovOik_UniAx:2019,OdintsovOik_Axion:2020,Oikonomou_Uni:2021,OikonomouFTR:2023,OdintsovSGS_Axi:2023}:
 \begin{equation}
 V(\phi)=\frac12 m_a^2 \phi^2\ .
 \label{Vphi}\end{equation}
Moreover, the modified FLRW equations (\ref{eqn1}), (\ref{eqn2}) might be expressed a a dynamical system by using the
relation $R=6\dot H + 12H^2$ \cite{OdintsovSGS:2017,OdintsovSGSlog:2019,OdintsovSGS_Axi:2023},
 \begin{eqnarray}
\frac{dH}{d\log a}&=&\frac{R}{6H}-2H, \label{dynam1}\ , \\
\frac{dR}{d\log
a}&=&\frac1{F_{RR}}\bigg(\frac{\kappa^2\rho_\mathrm{tot}}{3H^2}-F_R+\frac{RF_R-F}{6H^2}\bigg)\ .
 \label{dynam2}
 \end{eqnarray}
 Note that the continuity equations (\ref{cont} for matter and for thee axion (\ref{eqphi}) are a consequence of this pair of equations, such that they are not independent. One should also note that the $F(R)$ model  (\ref{FR2}) becomes the usual exponential $F(R)$ gravity model in the limit $\alpha=0$ (see  Refs.~\cite{Linder2009,nostriexp,OdintsovSGS:2017}) with no logarithmic corrections, which recovers the usual $\Lambda$CDM at $\beta\to+\infty$ or $R\to+\infty$. However, for $\alpha\neq 0$ the Lagrangian (\ref{FR2}) does not lead to $\Lambda$CDM model at any limit, since  the logarithmic term remains:
 $$
 F(R)\simeq R-2\Lambda\left( 1-\alpha{\cal R}\log\frac{\cal R}2\right),\qquad\mbox{if}\qquad \beta{\cal R}\gg 1.
 $$
On the other hand, any $F(R)$ model has to satisfy some viability conditions in order to turn out a consistent theory that recovers GR predictions at local scales. In particular, the effective gravitational constant at high curvature regimes must remain positive such that the following relation should hold during the post-inflationary era \cite{OdintsovSGSlog:2019}:
 \begin{equation}
\left|F_R(R)-1\right|\ll 1 \quad \Rightarrow \quad
\alpha\left(1+\log\frac{R}{4\Lambda}\right)\ll 1\,.
 \label{alpll1}\end{equation}
 As a consequence, $\alpha$ in the model (\ref{FR2}) must satisfy the condition $\alpha\ll1$. As shown below, such a condition is fulfilled for the best fit values for this parameter.

\section{Cosmological evolution}
\label{Late}

In this section the evolution of the Hubble parameter $H$ and the Ricci
scalar $R$ (as well as the axion $\phi$ for the particular case) are analysed, focusing on the relevant cosmological period concerning the data that is used later to test the models. In this sense, the earliest data comes from the Cosmic Microwave Background radiation (CMB) that refers to the photon-decoupling
epoch at redshifts $z\simeq1100$. Other observations (SNe Ia, CC, BAO) are provided for redshifts
$z\le2.4$. Hence, as the concerning period establishes redshifts $z<10^6$, which corresponds to ${\cal
R}<10^{18}$, the inflationary term can be neglected in \eqref{FR2}, since the normalized Ricci scalar at the end of inflation is many orders of magnitude larger, ${\cal R}_0\sim 10^{85}$.

Then, for redshifts $z<10^6$ one has to consider both a pressureless (non-relativistic) fluid as relativistic
particles (radiation). In the first scenario, $\rho_m$ contains baryons and dark matter:
$\rho_m= \rho_b+\rho_{dm}$ while radiation energy density is given by $\rho_r$. The cosmological evolution for these components is computed by the
continuity equation (\ref{cont}) and equations (\ref{rhoa})\,--\,(\ref{Vphi}) for axions in the second scenario:
  \begin{equation}
 \rho_\mathrm{tot}=\left\{\begin{array}{ll}
 \rho_m^0a^{-3}+ \rho_r^0a^{-4}, \; &\text{Case 1:}\;\,\rho_m= \rho_b+\rho_{dm}\,;\\
 \rho_b^0a^{-3}+ \rho_r^0a^{-4}+\frac12(\dot\phi^2+m_a^2 \phi^2),&\text{Case 2:}\ \rho_{dm}=\rho_{\phi}.
 \end{array}\right.
 \label{rho}\end{equation}
 where $\rho_m^0$, $\rho_b^0$  and $\rho_r^0$ are the energy densities for dust, baryons and radiation at the
present time $t_0$, respectively. In order to reduce the number of free parameters, one can fix below the radiation-matter and baryon-matter ratios as provided by Planck
\cite{Planck13,Planck18,OdintsovSGS_Axi:2023}:
   \begin{equation}
X_r=\frac{\rho_r^0}{\rho_m^0}=2.9656\cdot10^{-4}, \qquad
X_b=\frac{\rho_b^0}{\rho_m^0}\simeq0.1574\\ . \label{Xrm}
 \end{equation}

Despite the exponential $F(R)$ model with logarithmic terms might deviate from 
$\Lambda$CDM model at high redshifts, as shown above, under the
condition (\ref{alpll1}) or for  $\alpha\ll1$, the Hubble parameter and the Ricci scalar might be close to the $\Lambda$CDM asymptotic expressions given by 
\cite{OdintsovSGSlog:2019,OdintsovSGS_Axi:2023}:
 \begin{equation}
 H^2=H^{*2}_0\Big[\Omega_m^{*} \big(a^{-3}+ X_r a^{-4}\big)+\Omega_\Lambda^{*}\Big]\ ,\qquad
 \frac{R}{2\Lambda}=2+\frac{\Omega_m^{*}}{2\Omega_\Lambda^{*}}a^{-3}\ .
  \label{asymLCDM}\end{equation}
Here $H^{*}_0$ is the Hubble parameter as measured today whether the underlying evolution would be described by the $\Lambda$CDM model. In general, this value will differ from the Hubble constant $H_0=H(t_0)$ as predicted for the concerning $F(R)$ gravity model, even under the assumption that the modified gravity model matches  (\ref{asymLCDM}) at redshifts $10^3\le z\le10^5$, since the posterior evolution will deviate from the one of the $\Lambda$CDM scenario. This also concerns the difference between density parameters $\Omega_i^*$ and
$\Omega_i^0=\kappa^2\rho_i(t_0)/(3H_0^2)$. However these parameters are related in the following way
\cite{OdintsovSGS:2017,OdintsovSGSlog:2019,OdintsovSGS_Axi:2023}:
 \begin{equation}
 \Omega_m^0H_0^2=\Omega_m^{*}(H^{*}_0)^2=\frac{\kappa^2}3\rho_m(t_0)\ ,
 \qquad  \Omega_\Lambda H_0^2=\Omega_\Lambda^{*}(H^{*}_0)^2=\frac{\Lambda}3\ .
  \label{H0Omm}\end{equation}
It is also convenient to use $H^*_0$ for normalising the Hubble parameter to express the expansion ratio as follows
\begin{equation}
E=\frac{H}{H_0^{*}}\ .
   \label{E}\end{equation}
By using this expression, the dynamical equations (\ref{dynam1}) and (\ref{dynam2}) can be rewritten with the dimensionless variables  $E(a)$, ${\cal R}(a)$
\cite{OdintsovSGSlog:2019} as:
\begin{eqnarray}
\frac{dE}{dx}&=&\Omega_\Lambda^{*}\frac{{\cal R}}{E}-2E,\qquad x=\log a\ , \label{eqE} \\   
\frac{d\log{\cal R}}{dx}&=&\frac{\big\{ E^2_m+\Omega_\Lambda^*\big[\alpha{\cal R}
 \big(1-\epsilon+\beta\epsilon{\cal R}\ell)\big)-\epsilon(1+\beta {\cal R})\big] \big\}\big/E^2-1
 +\beta \epsilon -\alpha\Pi}
 {\alpha+\alpha\,\epsilon\big[-1+\beta {\cal R}(2+2\ell-\beta {\cal R}\ell ) \big]
 +\beta^2\epsilon {\cal R}}\ ,
  \label{eqR}
  \end{eqnarray}
  where $\epsilon=e^{-\beta {\cal R}}$, $\ell=\log({\cal R}/2)$, $\Pi=1+\ell-\epsilon\big[1+(1-\beta {\cal R})\ell \big]$
  and
 \begin{equation}
E^2_{m}=
 \left\{\begin{array}{ll}
 \Omega_m^{*}(a^{-3}+ X_r a^{-4})+\Omega_\Lambda^{*},& \;1:  \mbox{ \ without axion}\,,\\
 \Omega_m^{*}(X_ba^{-3}+ X_ra^{-4})+\frac16(\Psi^2+\mu_a^2\Phi^2)+\Omega_\Lambda^*,&\;2:
 \mbox{ \ with axion}\,.
 \end{array}\right.
 \label{Emat}\end{equation}
Moreover, for the case involving the axion field, the following dimensionless variables are defined:
\begin{equation}
\Phi=\kappa\phi,\qquad \Psi=\frac{\kappa\dot\phi}{H_0^{*}},\qquad
\mu_a=\frac{m_a}{H_0^*}\,.
   \label{Phimu}\end{equation}

\begin{figure}[ht]
   \centerline{ \includegraphics[scale=0.7,trim=5mm 0mm -2mm 0mm]{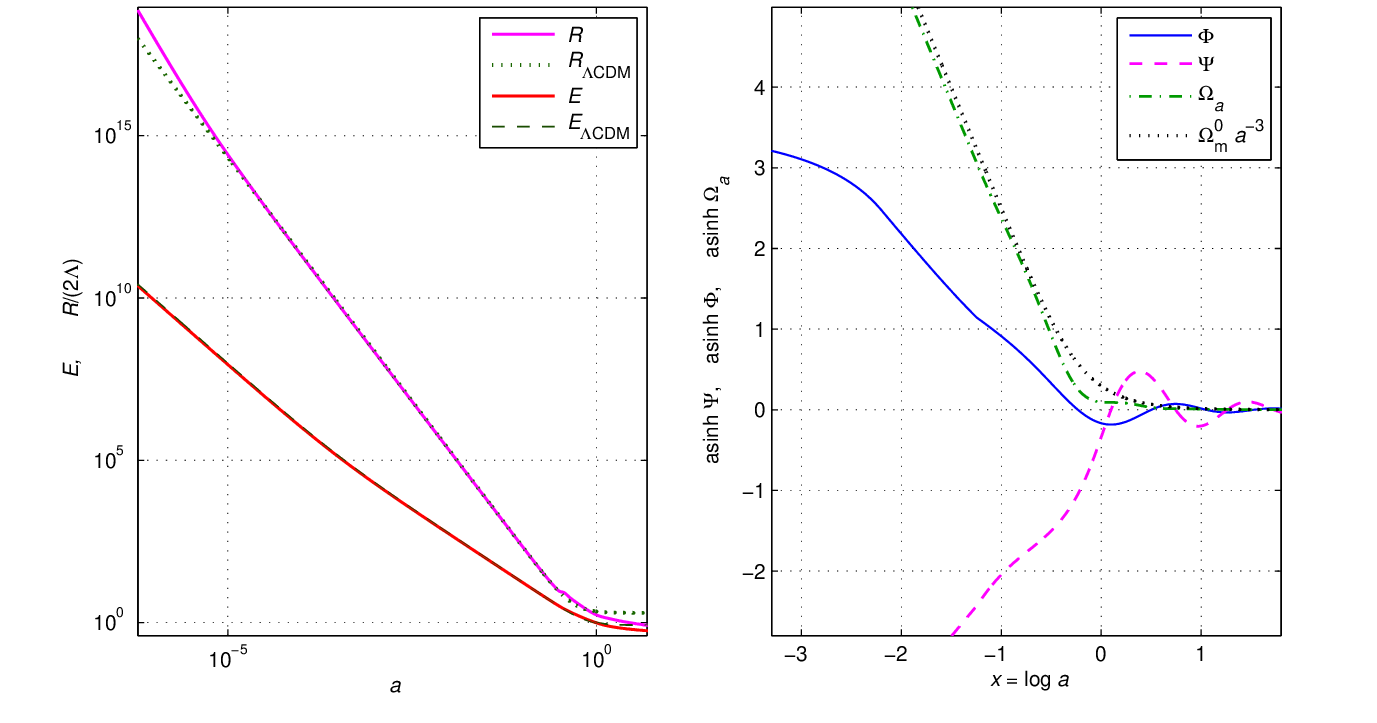}}
\caption{Evolution of the normalised Hubble parameter $E(a)$, the Ricci scalar ${\cal
R}(a)$ (left panel), the axion amplitudes $\Phi(x)$, $\Psi(x)$ and $\Omega_a(x)$ for
the $F(R)$ model (\ref{FR2})  in comparison with the $\Lambda$CDM model
(\ref{asymLCDM}). }
  \label{F1}
\end{figure}

For the first scenario (\ref{FR2}) with dust-like dark matter we can solve numerically
the system (\ref{eqE})-(\ref{eqR}) starting by some appropriate initial conditions at $a_\mathrm{ini}=e^{x_\mathrm{ini}}$, where some restrictions must be satisfied
\cite{OdintsovSGSlog:2019}. In this sense, the starting point for the integration must lie later than the
inflationary era but before the recombination epoch ($a_\mathrm{ini}<10^{-3}$), while the
factor $\epsilon=e^{-\beta {\cal R}}$ at this point should be small
$\epsilon(a_\mathrm{ini})\ll 1$, which means that $\beta {\cal R}_\mathrm{ini}\gg 1$. For the cases depicted in Fig.~\ref{F1} as well as for further calculations the scale factor $a_\mathrm{ini}\simeq10^{-10}$ is assumed (remind that $a(t_0)=1$). \\

In the vicinity of $a_\mathrm{ini}$ the Universe goes through the radiation dominated
epoch, such that the solutions  of the system (\ref{eqE})-(\ref{eqR}) are assumed to behave as:
\begin{equation}
{\cal R}\simeq {\cal A} a^{-4}={\cal A} e^{-4x},\qquad E^2\simeq {\cal B} a^{-4}\ .
   \label{RElim}\end{equation}
Here ${\cal A}$ and ${\cal B}$ are two positive constants to be determined by substituting the expressions (\ref{RElim}) into the equations (\ref{eqE})-(\ref{eqR}) for
the limit $\beta {\cal R}\gg 1$, leading to the following relations:
\cite{OdintsovSGSlog:2019}:
\be
2\alpha(\Omega_m^* X_r+\alpha\Omega_\Lambda^*{\cal A})=\Omega_\Lambda^*{\cal
A}\Xi^2\ ,\;\quad \Xi=1-\alpha\Big(4x_\mathrm{ini}+3-\log\frac{\cal A}2\Big)\ ,\;\quad
 {\cal B}=(\Omega_m^* X_r+\alpha\Omega_\Lambda^*{\cal A})/\Xi\ .
 \ee
These equations provide the asymptotical amplitudes ${\cal A}$ and ${\cal B}$ in the
expressions (\ref{RElim}), which can be used as initial conditions in $E(x_\mathrm{ini})$,
${\cal R}(x_\mathrm{ini})$ for the equations (\ref{eqE})-(\ref{eqR}). In the first scenario, this system is integrated numerically at $x\ge x_\mathrm{ini}$ as described in
Ref.~\cite{OdintsovSGSlog:2019}.

For the second scenario with the axion field $\phi$ one has to add  the dynamical equation (\ref{eqphi}) for  $\phi$ to the set of Eqs.~(\ref{eqE})-(\ref{eqR}), which by following the notation
(\ref{Phimu}) may be reduced to the system \cite{OdintsovSGS_Axi:2023}
 \begin{equation}
\frac{d\Phi}{dx}=\frac1{E}\Psi,\qquad \frac{d\Psi}{dx}=-3\Psi-\frac{\mu_a^2}{E}\Phi\ .
 \label{eqphi2}
  \end{equation}
Initial conditions for the axion field $\Phi$ and its derivative $\Psi$ can be determined by using the
approach provided in Ref.~\cite{OdintsovSGS_Axi:2023}, where one assumes that the axion energy
density (\ref{rhoa}) behaves like cold dark matter, i.e. $\rho_a\simeq\rho_a^0a^{-3}$ during the epoch when the Hubble parameter
$H(a)$ is of the order of the axion mass $m_a$ (whereas $\phi^2$ is frozen when $H\gg m_a$)
\cite{OdintsovOik_UniAx:2019,OdintsovOik_Axion:2020,Oikonomou_Uni:2021,OikonomouFTR:2023}.\\

Hence, one supposes that during the epoch when $H(a)$ is of the order of $m_a$, the cosmological evolution $E(a)\sim\mu_a$ and the axion energy density  $\rho_a$ mimics dark matter:
\begin{equation}
\rho_a=\rho_{dm}=\rho_{dm}^0a^{-3},\qquad\rho_{dm}^0=\rho_{m}^0-\rho_{b}^0\ ,\qquad
 E(a)\sim\mu_a\ .
 \label{rhoadm}
 \end{equation}
 Then, at the moment' $x=x_*$ when $E(a)=\mu_a$, the axion field makes a smooth transition from a pressureless fluid (\ref{rhoadm}) to a dynamical equation of state that is determined by the equations  (\ref{eqE})-(\ref{eqR}) and (\ref{eqphi2}). These conditions of smoothness at the point $x=x_*$ yields:
 \begin{equation}
\rho_a(x_*)=\rho_{dm}^0e^{-3x_*}\ ,\qquad
\frac{d\rho_a}{dx}\Big|_{x_*}=-3\rho_{dm}^0e^{-3x_*}\ ,
 \label{smooth}
 \end{equation}
 whereas the functions $\Phi$ and $\Psi$ in (\ref{Phimu})  lead to the relations  \cite{OdintsovSGS_Axi:2023}
 \begin{equation}
\Psi^2\big|_{x_*}=\mu_a^2\Phi^2\big|_{x_*}=3\Omega_m^*(1-X_b)\,e^{-3x_i*}\ ,
   \label{Psiini}\end{equation}
 These expressions are assumed to be the initial conditions for $\Phi$ and $\Psi$ in the equation
(\ref{eqphi2}) at the transition point $x=x_*$ (recall that the variable $x=\log a$ refers to the number of e-folds).

Fig.~\ref{F1} depicts the cosmological evolution for a particular case after solving the system (\ref{eqE})-(\ref{eqR}) and (\ref{eqphi2}) with
$\Omega_m^*=0.3$, $\Omega_\Lambda^*=0.7$, $\alpha=0.005$, $\beta=1$, $\mu_a=3$.

In the left panel of Fig.~\ref{F1} one can see that the normalised Hubble parameter
$E(a)$ for the  $F(R)$ model behaves similarly to $\Lambda$CDM model,
deviating just at current times. However, the corresponding plots for ${\cal R}(a)$
diverge not only at late times, but also at $a<10^{-5}$ ($z>10^{5}$) due to the presence of the
logarithmic corrections in the Lagrangian (\ref{FR2}). In the right panel of Fig.~\ref{F1}, the axion amplitudes are shown as well as the energy density parameter for the axion field
and is compared with the matter
density parameter $\Omega_m^*$. These functions grow rapidly as $a$ decreases, so inverse hyperbolic functions (asinh$\,x=\log(x+\sqrt{1+x^2})$ are used instead. The
 amplitudes $\Phi(x)$ and $\Psi(x)$ oscillate after crossing the transition point $x=x_*$.\\

In the following section, the above scenarios (\ref{FR2})  are confronted with the
observational data and compare the results with standard exponential gravity (in absence of logarithmic corrections) and the $\Lambda$CDM model.

\section{Testing the models with observational data}
\label{Results}

Let us now test the two scenarios described above within the  $F(R)$ model (\ref{FR2}) and compare
its observational predictions with data from Type Ia supernovae (SNe Ia), baryon acoustic oscillations (BAO),  estimations of the Hubble
parameter $H(z)$ or Cosmic Chronometers (CC)  and parameters from the cosmic microwave
background radiation (CMB). For this purpose the $\chi^2$ function is computed, where
the contributions from  SNe Ia, CC $H(z)$ data, CMB  and BAO are included:
 \begin{equation}
  \chi^2=\chi^2_\mathrm{SN}+\chi^2_H+\chi^2_\mathrm{CMB}+\chi^2_\mathrm{BAO}\ .
 \label{chitot} \end{equation}
In order to calculate the contributions $\chi^2_j$ from the four sources of data the methods and approaches suggested earlier in the paper
\cite{OdintsovSGS_Axi:2023} are followed, which are briefly described in Appendix \ref{App}.

In the first scenario when considering dark matter just as a pressureless fluid $\rho_m=\rho_{m}^0a^{-3}$), the radiation-matter and baryonic-matter ratios (\ref{Xrm}) are fixed, such that 5 free
parameters remain for this model:
 \begin{equation}
 \alpha, \quad\beta, \quad \Omega_m^0, \quad \Omega_\Lambda,\quad H_0\,.
\label{FreeParam} \end{equation}
 For the second scenario when assuming dark matter to be an axion field, another free parameter arises, $\mu_a$, in (\ref{Phimu}). \\

Note also that we make calculations for the parameters  $\Omega_m^0$, $\Omega_\Lambda$, $H_0$ instead of $\Omega_m^*$,
$\Omega_\Lambda^*$, $H_0^*$ by using the relations (\ref{H0Omm}) between both sets of parameters.

\begin{figure}[th]
   \centerline{ \includegraphics[scale=0.64,trim=5mm 0mm 2mm -1mm]{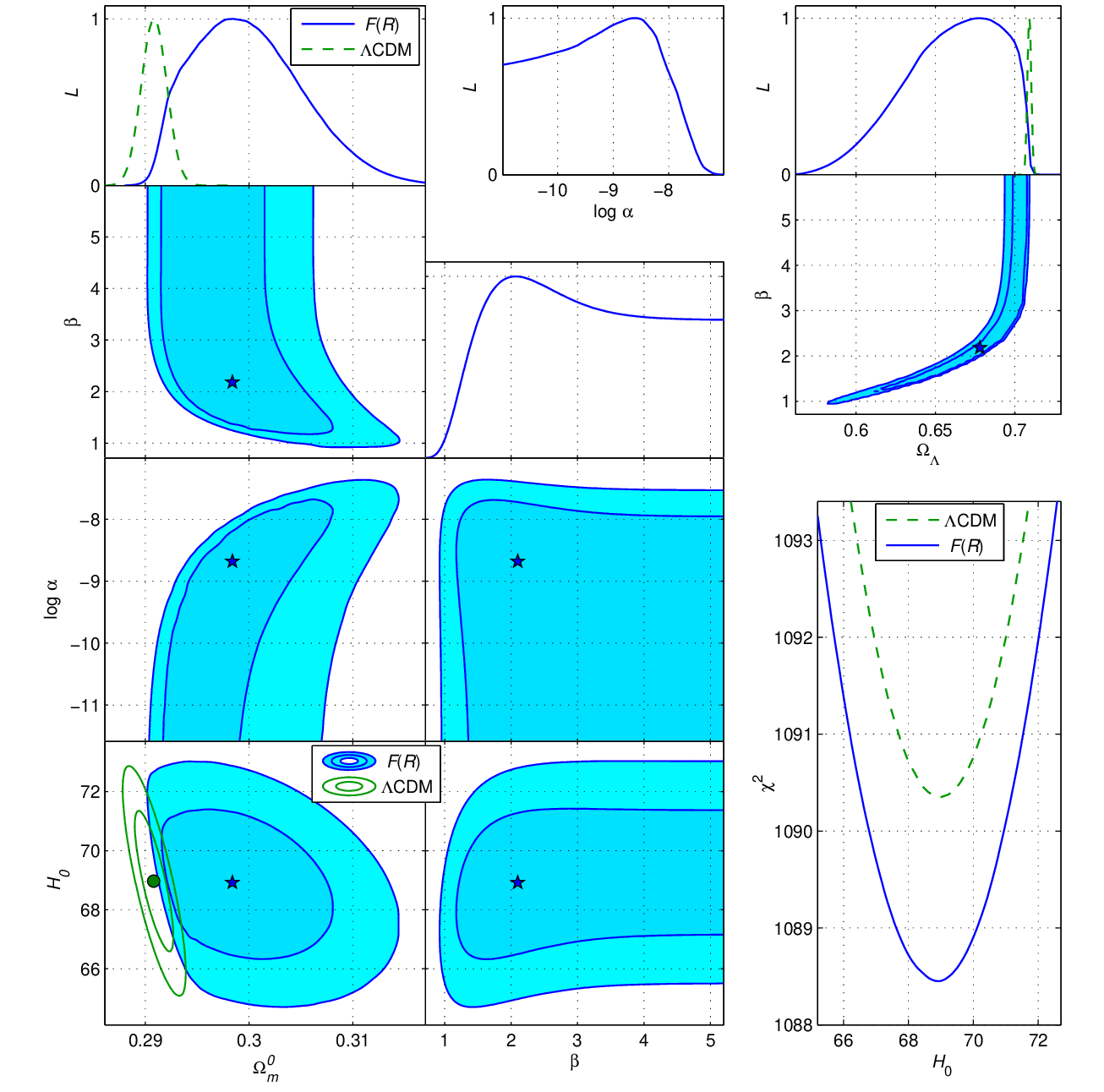}}
\caption{Contour plots of $\chi^2$ with $1\sigma$, $2\sigma$ CL likelihood functions $
{\cal L}(\theta_i)$ and one-parameter distributions $\chi^2(H_0)$ for the  exponential
$F(R)$ model  (\ref{FR2}) without axion in comparison with the $\Lambda$CDM model. }
  \label{F2}
\end{figure}

One should keep in mind that in the limit $\alpha=0$ (without  logarithmic corrections)
this model turns out the standard exponential $F(R)$ case proposed in Ref.~\cite{Linder2009} whereas the $\Lambda$CDM scenario is recovered for $\alpha=0$ and $\beta\to+\infty$. In order to
fit the model with observational data and calculate the best fit values of the free parameters
(\ref{FreeParam}), the technique of the maximum likelihood is followed and the corresponding
2 parameters contour plots are obtained by marginalising over the rest of the free parameters, as shown in Fig.~\ref{F2}. The contours correspond to
$1\sigma$ (68.27\%) and $2\sigma$ (95.45\%) confidence regions for the two-parameter
distributions $\chi^2(\theta_i,\theta_j)$ calculated via minimization of $\chi^2$ over all the remaining free parameters. For example, in the bottom-left panel of
Fig.~\ref{F2} the contours for
 $\chi^2(\Omega_m^0,H_0)=\min\limits_{\alpha,\beta,\Omega_\Lambda}\chi^2(\alpha,\dots,H_0)$ are depicted.

The blue stars for the $F(R)$ model and the green circles for $\Lambda$CDM denote the
best fits of the corresponding two-dimensional distributions. These best fits  together with the $1\sigma$
error for the free model parameters (\ref{FreeParam})  are summarised in
Table~\ref{Estim}. The $1\sigma$ errors  are calculated via one-parameter
distributions $\chi^2(\theta_j)$  and likelihoods ${\cal L}(\theta_j)$:
   \begin{equation}
\chi^2_{tot}(\theta_j)=\min\limits_{\mbox{\scriptsize other
}\theta_k}\chi^2_(\theta_1,\dots),\qquad {\cal L}_{tot}(\theta_j)= \exp\bigg[-
\frac{\chi^2(\theta_j)-m^\mathrm{abs}}2\bigg]\ ,
 \label{likeli} \end{equation}
 where $\theta_j$ is the model parameter, being $m^\mathrm{abs}$ the absolute minimum for $\chi^2$.

The best fit for $\alpha$ is small
enough to satisfy the condition (\ref{alpll1}). Hence, for convenience, in the panels of Fig.~\ref{F2}, a logarithmic scale is used along the axes. The one-parameter distribution $\chi^2(H_0)$ is shown in the bottom-right panel of
Fig.~\ref{F2} in comparison with the one from $\Lambda$CDM model
 \begin{equation}
 H^2=H_0^2\big[\Omega_m^0 (a^{-3}+ X_r a^{-4})+\Omega_\Lambda\big]\, ,\qquad
 \Omega_\Lambda=1-\Omega_m^0\ .
  \label{LCDM}\end{equation}
The absolute minimum  $m^\mathrm{abs}=\min\chi^2\simeq1088.45$ for the $F(R)$ model
(\ref{FR2}) with logarithmic corrections is essentially better than the $\Lambda$CDM
result $1090.35$. This difference supports the results of the paper
\cite{OdintsovSGSlog:2019}, where other observational data were considered. Note that the
standard  exponential $F(R)$ model without the logarithmic term in
Ref.~\cite{OdintsovSGS_Axi:2023} leads to the minimum $m^\mathrm{abs}=1090.21$ (see also
Table~\ref{Estim}) that is only a bit better than the $\Lambda$CDM value.\\

For the matter density parameter $\Omega_m^0$ in the $F(R)$ model, the best fit is larger and the contours are wider than for
the $\Lambda$CDM model, as shown up in Table~\ref{Estim}, and may be connected with
the additional degrees of freedom within the  $F(R)$ model (\ref{FR2}) for a finite value of
$\beta$  in comparison with the two free parameters of the $\Lambda$CDM
model (\ref{LCDM}). However, the best fits of the Hubble constant  for both 
models are rather close.\\

\begin{figure}[th]
  \centerline{\includegraphics[scale=0.71,trim=4mm 4mm 5mm 5mm]{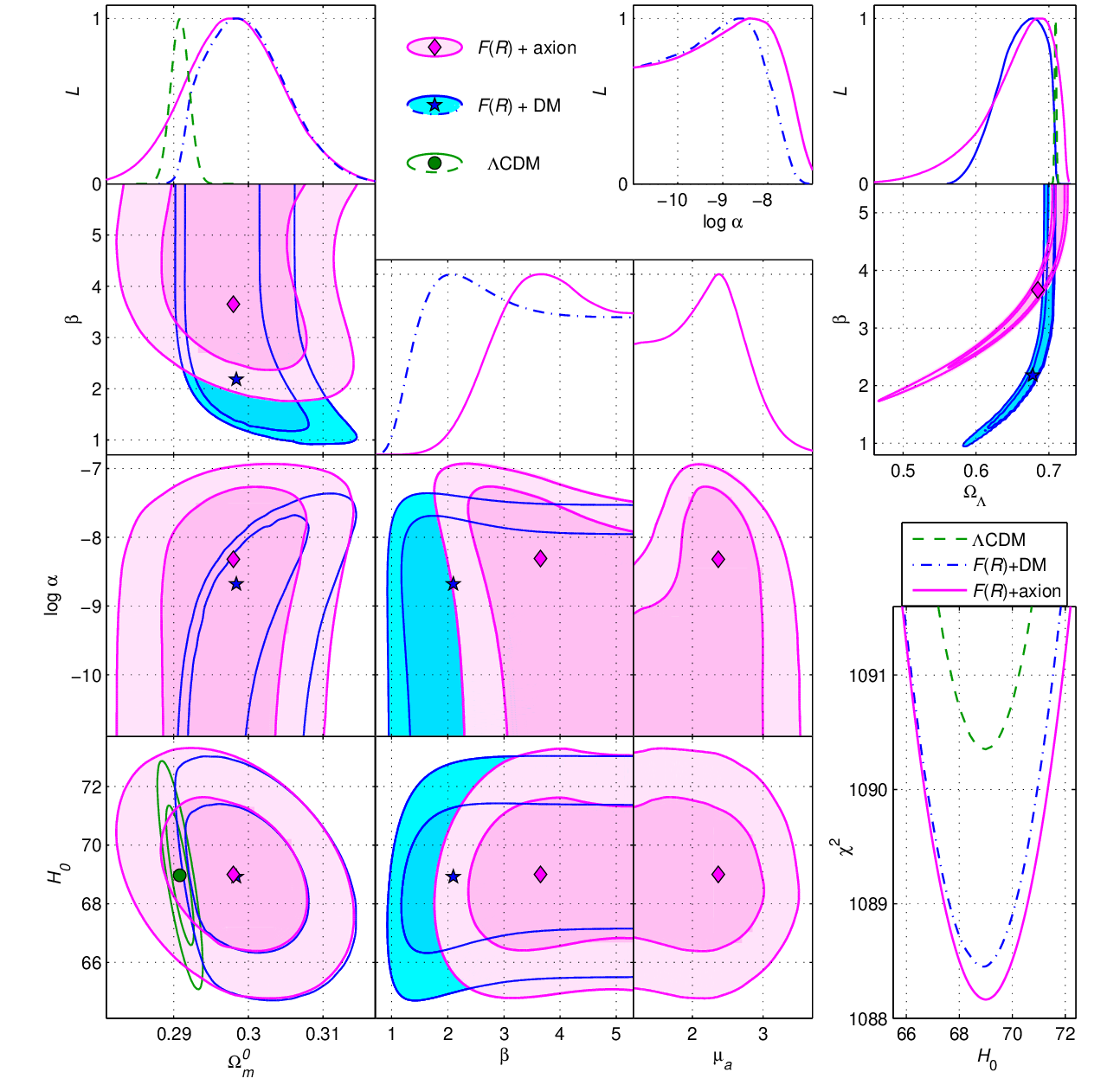}}
\caption{The second $F(R)$ scenario (\ref{FR2}) with the axion field:  contour plots of
$\chi^2$ with $1\sigma$, $2\sigma$ CL and likelihoods in comparison with the first
scenario without axion and the $\Lambda$CDM model. }
  \label{F3}
\end{figure}

For the second $F(R)$ scenario (\ref{FR2}) with the presence of the axion field, calculations turn out a bit more
complicated for the space of six free parameters. By solving the system of equations
(\ref{eqE})-(\ref{eqR}) and (\ref{eqphi2}) with the above initial conditions
(\ref{asymLCDM}) at $x_\mathrm{ini}$ and (\ref{Psiini}) at the transition point $x=x_*$, when $E(x)=\mu_a$. Results of the calculations for this scenario are shown in Fig.~\ref{F3}, where it is compared 
with the previous scenario and the $\Lambda$CDM model by following the same notation as in
Fig.~\ref{F2}.

As shown in the bottom-right panel in Fig.~\ref{F3}, the one-parameter distribution
$\chi^2(H_0)$ for all the models show similar shapes and the axion case reaches the best minimum, given by
$m^\mathrm{abs}=1088.16$, in comparison with the other models. This result can be connected with the additional free parameter $\mu_a$,  while best fit $\beta=3.23^{+\infty}_{-1.17}$ is slightly larger than in the previous case, which can explain the small difference of $m^\mathrm{abs}$ with the first scenario. In the bottom-left panel of Fig.~\ref{F3}, the $1\sigma$ and $2\sigma$ CL
domains for the axion $F(R)$ model with the same contours of the other models are shown for the $\Omega_m^0-H_0$ plane. Both $F(R)$
scenarios predict very close best fits for  $\Omega_m^0$ and $H_0$ but with larger errors, as shown by the  $1\sigma$ and $2\sigma$ CL domains along the $\Omega_m^0$ axis.

The best fit for $\alpha$ in the axion scenario is a bit larger than the one of the first case without axion. However both values satisfy the condition $\alpha\ll1$.

One can see in Table \ref{Estim} that both cases of the $F(R)$ model (\ref{FR2})
show essentially better results for the absolute minimum
$m^\mathrm{abs}=\min\chi^2$ with respect to the the standard exponential $F(R)$ model
\cite{OdintsovSGS_Axi:2023} and the $\Lambda$CDM model.  However, $\Lambda$CDM model
still presents a better goodness of the fits when accounting for the number of
free model parameters  $N_p$ and follow the Akaike information:
criterion \cite{Akaike}
 \be
 \mbox{AIC} = \min\chi^2_{tot} +2N_p.
  \ee 
 The more free parameters of the model, the better fits data at the price of a larger Akaike parameter. For instance, the axion $F(R)$ model has
the largest number of parameters $N_p=6$ and also the largest AIC, as shown in Table \ref{Estim}.

 {\color{red} 
\begin{table}[ht]
\begin{tabular}{|l|c|c|c|c|c|c|c|c|}
\hline  Model &   $\min\chi^2/d.o.f$& AIC & $\Omega_m^0$& $\Omega_\Lambda$& $H_0$ & $\beta$ & $\alpha\cdot10^4$ & $\mu_a$\\
\hline
  Exp+log\,$F(R)$ & 1088.45 /1106 & 1098.45& $0.2984^{+0.0064}_{-0.0057}$ & $0.679^{+0.026}_{-0.044}$
& $68.92^{+1.63}_{-1.72}$& $2.06^{+\infty}_{-0.68}$ & $1.84^{+1.43}_{-1.77}$  & - \rule{0pt}{1.1em}  \\
\hline
 Exp+log\,$F(R)$+axion & 1088.16 /1105 & 1100.16& $0.2980^{+0.0065}_{-0.0065}$&$0.685^{+0.031}_{-0.049}$ & $69.0^{+1.72}_{-1.71}$& $3.66^{+\infty}_{-0.92}$ & $2.44^{+3.06}_{-2.42}$ & $2.37^{+0.39}_{-1.24}$ \rule{0pt}{1.1em}\\
\hline
Exp $F(R)$ & 1090.21 /1108 & 1098.21& $0.2913^{+0.0035}_{-0.0015}$ & $0.703^{+0.007}_{-0.047}$ & $68.84^{+1.75}_{-1.64}$& $2.94^{+\infty}_{-1.325}$&-&- \rule{0pt}{1.1em}  \\
\hline
$\Lambda$CDM& 1090.35 /1110 & 1094.35& $0.2908^{+0.0013}_{-0.0012}$&$0.7092^{+0.0012}_{-0.0013}$& $68.98^{+1.58}_{-1.60}$ & -& -& - \rule{0pt}{1.1em}  \\
\hline
 \end{tabular}
 \caption{Summary of the best fit values for the free parameters and $\min\chi^2$ for the two dark matter models within exponential $F(R)$ gravity with logarithmic corrections  in comparison with the standard exponential $F(R)$ model \cite{OdintsovSGS_Axi:2023}
and $\Lambda$CDM model.}
\label{Estim}
\end{table}
}

%
\section{Conclusions}
\label{conclusions}
%

A particular type of $F(R)$ gravity has been considered, specifically the paper has focused on a model that includes a negative exponential of the curvature scalar and some logarithmic terms. Exponential $F(R)$ gravity has been widely analysed in the literature, where shown that can reproduce well the cosmological evolution and at the same time it is capable of reproducing GR predictions at local scales \cite{Linder2009,OdintsovSGS:2017}. Logarithmic terms are considered in the gravitational action as they provide a smooth transition from one cosmological epoch to another and neither introduces large corrections that might violate local gravity tests. Then, two types of dark matter have been assumed. Firstly, the usual description of dark matter as an effective pressureless fluid and secondly, as an axion field. The latter is well-known in the analysis of quantum field theories and the CP violation in quantum chromodynamics and for a particular range of masses is known that might contribute or constitute completely the amount of dark matter of the universe. Hence, the analysis of both models becomes essential to find out ways to break the great degeneracy on the large number of cosmological models that reproduce the universe evolution in a correct way. \\

For this analysis, a comparison with the latest observational data is followed. To do so, data from Sne Ia, Hubble parameter data, CMB and BAO is used and the exponential $F(R)$ gravity with logarithmic terms is compared for the two possible dark matter species, where the fits of the free parameters are computed by using the technique of the minimum $\chi^2$. Fits for the $\Lambda$CDM model and exponential gravity without logarithmic corrections are also performed. The results, summarised in Table \ref{Estim} and also shown in Figs.~\ref{F2}-\ref{F3}, suggest that while the best fit provides a smaller value for $\chi^2_{min}$ for both cases of exponential $F(R)$ gravity with logarithmic terms, the goodness of the fits, computed by the AIC coefficient, favour $\Lambda$CDM model and exponential $F(R)$ gravity in absence of logarithmic corrections. The point lies on the larger number of free parameters for the two cases of the paper. However, a straight conclusion that is followed regarding the presence of the axion field is that data supports better the description of dark matter as a pressureless fluid. Nevertheless, statistically speaking, one can not rule out such type of axion models by estimating the best fits for the free parameters with the available cosmological data. Moreover, the presence of logarithmic terms may not be favoured in comparison to models in absence of such a type of terms despite the difference on the goodness of the fits is not significant, but the reduction of extra terms and extra free parameters point out to a way of reducing the complexity of the effective modifications of GR. \\

Hence, despite this detailed analysis, one can not conclude that a particular model fits better the observational data. However, this type of modifications of GR seem to behave well at all levels, such that they point out to the way that might be followed for extending GR in order to reach a more complete theory of gravity. Additional analysis regarding other frameworks, as compact objects or gravitational waves, should be followed in order to determine the real physical possibilities of this type of modifications. Finally, axion field remains as a good candidate for dark matter but a deeper analysis, including the growth of cosmological perturbations and gravitational instability al the non-linear regime, should be performed to know better the behaviour of this class of dark matter candidates.

\section*{Acknowledgments}
This work was  supported by MICINN (Spain) projects PID2019-104397GB-I00 (S.D.O.) and PID2020-117301GA-I00 (D.S.G.) funded by MCIN/AEI/10.13039/501100011033 (``ERDF A way of making Europe" and ``PGC Generaci\'on de Conocimiento") and also by the program Unidad de Excelencia Maria de Maeztu CEX2020-001058-M, Spain (S.D.O).

\section*{Appendix}
\label{App}
In this section the different sources of observational data are described, which include:
(a) Pantheon sample of Type Ia supernovae (SNe Ia) data \cite{Scolnic17}; (b)
measurements of the Hubble parameter $H(z)$ from Cosmic Chronometers (CC), (c) Cosmic
Microwave Background radiation (CMB) data and (d) Baryon Acoustic Oscillations (BAO). A
detail description of the corresponding
data analysis, methods, sources can be found in Refs.~\cite{OdintsovSGS_Axi:2023,OdintsovSGS:2022,OdintsovOS:2023}.\\

The SNe Ia contribution to the $\chi^2$ function (\ref{chitot}) is given by:
 $$ 
\chi^2_{\mathrm{SN}}(\theta_1,\dots)=\min\limits_{H_0} \sum_{i,j=1}^{N_\mathrm{SN}}
 \Delta\mu_i\big(C_{\mathrm{SN}}^{-1}\big)_{ij} \Delta\mu_j,\qquad
 \Delta\mu_i=\mu^\mathrm{th}(z_i,\theta_1,\dots)-\mu^\mathrm{obs}_i\ ,
 $$ 
 where  $N_{\mathrm{SN}}=1048$ datapoints of the distance moduli $\mu_i^\mathrm{obs}$ at redshifts
$z_i$ as provided by the Pantheon sample database \cite{Scolnic17}, while  $\theta_j$
are free model parameters, $C_{\mbox{\scriptsize SN}}$ is the covariance matrix and
$\mu^\mathrm{th}$ are the theoretical values:
 $$ 
 \mu^\mathrm{th}(z) = 5 \log_{10} \frac{(1+z)\,D_M(z)}{10\mbox{pc}},\qquad D_M(z)= c \int\limits_0^z\frac{d\tilde z}{H(\tilde
 z)}.
 $$ 

 For the Hubble parameter data  $H(z)$ we use here $N_H=32$
datapoints of Cosmic Chronometers (CC) given in Refs.~\cite{HzData}, i.e. measured as $H
(z)= \frac{\dot{a}}{a} \simeq -\frac{1}{1+z} \frac{\Delta z}{\Delta t}$ from different
ages $\Delta t$ of galaxies with close redshifts $\Delta z$. The corresponding $\chi^2$
function for CC $H(z)$ data yields:
 $$ 
    \chi_H^2(\theta_1,\dots)=\sum_{j=1}^{N_H}\bigg[\frac{H(z_j,\theta_1,\dots)-H^{obs}(z_j)}{\sigma _j}  \bigg]^2
 $$ 

For CMB data we use observational parameters  obtained from Planck 2018 data
\cite{Planck18} in the form \cite{ChenHuangW2018}:
 $$ 
  \mathbf{x}=\big(R,\ell_A,\omega_b\big),\qquad R=\sqrt{\Omega_m^0}\frac{H_0D_M(z_*)}c,\quad
 \ell_A=\frac{\pi D_M(z_*)}{r_s(z_*)},\quad\omega_b=\Omega_b^0h^2\ .
  $$ 
Here $z_*$ is the photon-decoupling redshift,
$h=H_0/[100\,\mbox{km}\mbox{s}^{-1}\mbox{Mpc}^{-1}]$,  $r_s(z)$ is the comoving sound
horizon. Details on the way to obtain $r_s(z)$ and other parameters are given in
Refs.~\cite{OdintsovSGS_Axi:2023,OdintsovSGS:2022,OdintsovOS:2023}. The corresponding
$\chi^2$ function for the CMB data is given by:
 $$ 
\chi^2_{\mbox{\scriptsize CMB}}=\min_{\omega_b}\Delta\mathbf{x}\cdot
C_{\mathrm{CMB}}^{-1}\big(\Delta\mathbf{x}\big)^{T},\qquad \Delta
\mathbf{x}=\mathbf{x}-\mathbf{x}^{Pl}
 $$ 
where the estimations
 $\mathbf{x}^{Pl}=\big(R^{Pl},\ell_A^{Pl},\omega_b^{Pl}\big)=\big(1.7428\pm0.0053,\;301.406\pm0.090,\;0.02259\pm0.00017\big)$
 and the covariance matrix $C_{\mathrm{CMB}}=\|\tilde C_{ij}\sigma_i\sigma_j\|$ are given in Ref.~\cite{ChenHuangW2018}.\\

For the baryon acoustic oscillations (BAO) data we consider two magnitudes:
 $$ 
d_z(z)= \frac{r_s(z_d)}{D_V(z)}\, ,\qquad A(z) =
\frac{H_0\sqrt{\Omega_m^0}}{cz}D_V(z)\,,
 $$ 
where $D_V(z)=\big[{cz D_M^2(z)}/{H(z)} \big]^{1/3}$, $z_d$ being the redshift at the
end of the baryon drag era. Here we use  21 BAO data points for $d_z(z)$ and 7 data
points for $A(z)$ from the papers tabulated in  Ref.~\cite{OdintsovSGS_Axi:2023} in the
following $\chi^2$ function:
 $$ 
\chi^2_{\mathrm{BAO}}(\Omega_m^0,\theta_1,\dots)=\Delta d\cdot C_d^{-1}(\Delta d)^T +
\Delta { A}\cdot C_A^{-1}(\Delta { A})^T\, .
 $$ 
Here, $\Delta d_i=d_z^\mathrm{obs}(z_i)-d_z^\mathrm{th}(z_i,\dots)$, $\Delta
A_i=A^\mathrm{obs}(z_i)-A^\mathrm{th}(z_i,\dots)$, $C_{d}$ and $C_{A}$ are the
covariance matrices for the correlated BAO data \cite{Percival:2009,Blake:2011}.

\end{document}